\documentclass[12pt]{iopart}

\usepackage{iopams}  
\usepackage{graphicx}
\expandafter\let\csname equation*\endcsname\relax
\expandafter\let\csname endequation*\endcsname\relax
\usepackage{amsmath}
\usepackage{color} 
\usepackage{hyperref}
\usepackage{babel}
\usepackage{textcomp,gensymb}
\usepackage{amsmath,bm}
\usepackage[font=small,labelfont=bf]{caption}
\hypersetup{
colorlinks=true,final=true,
        linkcolor=blue,
        citecolor=blue,
        filecolor=blue,
        urlcolor=blue,
}

\begin{document}
\title{First-principles investigation of the magnetoelectric properties of Ba$_7$Mn$_4$O$_{15}$}
\author{Urmimala Dey}
\address{Centre for Materials Physics, Durham University, South Road, Durham DH1 3LE, United Kingdom}
\author{Mark S. Senn}
\address{Department of Chemistry, University of Warwick, Gibbet Hill, Coventry CV4 7AL, United Kingdom}
\author{Nicholas C. Bristowe}
\address{Centre for Materials Physics, Durham University, South Road, Durham DH1 3LE, United Kingdom}
\ead{nicholas.bristowe@durham.ac.uk}
\vspace{10pt}

\begin{abstract}
 Type-II multiferroics, in which the magnetic order breaks inversion symmetry, are appealing for both fundamental and applied research due their intrinsic coupling between magnetic and electrical orders. Using first-principles calculations we study the ground state magnetic behaviour of Ba$_7$Mn$_4$O$_{15}$ which has been classified as a type-II multiferroic in recent experiments. Our constrained moment calculations with the proposed experimental magnetic structure shows the spontaneous emergence of a polar mode giving rise to an electrical polarisation comparable to other known type-II multiferroics. When the constraints on the magnetic moments are removed, the spins self-consistently relax into a canted antiferromagnetic ground state configuration where two magnetic modes transforming as distinct irreducible representations coexist. While the dominant magnetic mode matches well with the previous experimental observations, the second mode is found to possess a different character resulting in a non-polar ground state. Interestingly, the non-polar magnetic ground state exhibits a significantly strong linear magnetoelectric coupling comparable to the well-known multiferroic BiFeO$_3$, suggesting strategies to design new linear magnetoelectrics.
\end{abstract}

%
%
%
%
%

\section{Introduction}
The magnetoelectric (ME) effect refers to the coupling between the electrical and magnetic order parameters in a material allowing for an electrical control of magnetism or magnetic control of electrical polarisation~\cite{Schmid_1994,Fiebig_2005}. It has gained significant interest over the last couple decades due to various potential applications of ME materials in spintronics, magnetic sensors and non-volatile memory devices~\cite{Fiebig_2005,Fusil_2014,Wang_2014,Bauer_2012}. The mostly studied linear ME effect can be observed in polar and non-polar magnetic materials where the applied electric field induces weak ferromagnetic response via spin-orbit coupling or exchange striction mechanisms~\cite{Dzyaloshinskii_1959,Astrov_1960}.  Magnetoelectric multiferroic materials with multiple spontaneous ferroic orders in a single phase can exhibit general cross-coupling between electrical and magnetic order parameters and are thus ideal candidates to observe large ME effect~\cite{Scott_2006,Cheong_2007}. While type-I multiferroics, whereby the two ferroic orders have different unrelated origins, might be expected to display a weaker ME effect, type-II multiferroics are particularly interesting for their strong intrinsic ME coupling. Recently, Ba$_7$Mn$_4$O$_{15}$, a novel binary metal oxide, has been experimentally put forward as a possible type-II multiferroic where ferroelectricity was proposed to arise due to a distinct magnetic ordering of Mn$^{4+}$ ions which breaks spatial inversion symmetry~\cite{Clarke_2022}. 

The Sr$_7$Mn$_4$O$_{15}$ analogue, proposed by Kriegel et al.~\cite{Kriegel_1992,Kriegel_1999}, crystallises in a monoclinic phase containing face-sharing Mn$_2$O$_9$ octahedral dimer motifs. In 2017, Craddock et al. have assigned a canted antiferromagnetic ground state to Sr$_7$Mn$_4$O$_{15}$ which is attributed to the presence of a weak ferromagnetic ordering component~\cite{Craddock_2017}. However, recent symmetry-adapted refinement against neutron powder diffraction (NPD) data by Clarke et al. shows that the Mn ions inside the octahedral dimers exhibit strong antiferromagnetic coupling with the spins aligned along the monoclinic $b$-direction and there is no weak ferromagnetic component which can give rise to a multiferroic ground state~\cite{Clarke_2022}.

Magnetic structure refinements against NPD data for the isostructural Ba-variant, on the other hand, reveal the presence of a novel multiferroic ground state where the proposed noncentrosymmertic space groups are direct results of complex magnetic ordering~\cite{Clarke_2022}. In contrast to Sr$_7$Mn$_4$O$_{15}$, a binary combination of magnetic modes exists in Ba$_7$Mn$_4$O$_{15}$ which leads to canted antiferromagnetic spin structures at low temperature~\cite{Clarke_2022}. While polar displacements could not confidentially be assigned in earlier experiments, symmetry analyses of the magnetic structures that are consistent with the NPD data suggest two equally probable polar ground states with type-II multiferroic characters~\cite{Clarke_2022}. 

Here we perform density functional theory calculations to resolve the ground state magnetic behaviour of Ba$_7$Mn$_4$O$_{15}$. We first model the magnetic structure suggested in previous experiments by constraining the magnitudes and directions of the magnetic moments according to the experimental values~\cite{Clarke_2022}. Our constrained moment calculations reveal the spontaneous emergence of a non-zero polar mode associated with displacements of barium and oxygen atoms giving rise to a sizeable electric polarisation comparable to other known type-II multiferroics. We next remove the constraints on the magnetic moments and self-consistently relax the spins to obtain the true magnetic ground state from first-principles. Self-consistent calculations including spin-orbit coupling show that Ba$_7$Mn$_4$O$_{15}$ possesses a canted antiferromagnetic ground state where two magnetic modes transforming as distinct irreducible representations coexist. The dominant magnetic mode with $\sim$ 81\% contribution is found to be in excellent agreement with the previous experimental observations. However, the second contribution has a mode character different from the experimental findings leading to a non-polar ground state. Effects of different on-site Hubbard parameter $U$ and Hund’s coupling term $J$ have been investigated to confirm the stability of the ground state magnetic configuration. Although our self-consistent calculations do not predict a multiferroic ground state, magnetoelectric calculations show that Ba$_7$Mn$_4$O$_{15}$ displays a strong linear ME coupling. The largest component of the linear ME tensor is calculated to be three times the value predicted for the classic magnetoelectric Cr$_2$O$_3$~\cite{Iniguez_2008} and comparable to that of the well-known multiferroic BiFeO$_3$~\cite{Iniguez_2009}.   

\section{Computational Details}
In order to study the ground state magnetic behaviour of Ba$_7$Mn$_4$O$_{15}$, we employed the plane wave augmented (PAW) method within the density functional theory (DFT) framework as implemented in the Vienna Ab-initio Simulation Package (VASP)~\cite{VASP1,VASP2}, version 5.4.4. The PBEsol general gradient approximation (GGA)~\cite{PBEsol} to the exchange correlation functional was chosen for all the calculations to describe the equilibrium properties of our bulk oxide system accurately. We used the PAW pseudopotentials (PBE, version 5.4)~\cite{pseudo} with the following valence configurations: $5s^25p^66s^2$ (Ba), $3p^64s^23d^5$ (Mn) and $2s^22p^4$ (O). On-site correlation effects were considered within the framework introduced by Dudarev et al.~\cite{HubbardU}, where an effective on-site Hubbard parameter $U$ and Hund's parameter $J$ were applied on the $3d$ orbitals of Mn and varied for a range of values to check the stability of the ground state magnetic configuration. From convergence tests performed on a 52 atom unit cell, we found that a plane wave energy cutoff of 600 eV and a $3 \times 2 \times 2$ $k$-point mesh in the whole Brillouin zone (BZ) were sufficient to resolve the total energies, forces and stresses within 1 meV/formula unit, 1 meV/{\AA} and 0.001 GPa, respectively. An energy convergence criterion was set at $10^{-9}$ eV for all the calculations and full relaxations were performed until the Hellmann-Feynman forces on each atom were less than 0.1 meV/{\AA}. Spin-orbit coupling (SOC) effects were included self-consistently and a slightly denser $k$-mesh of $4 \times 3 \times 3$ was considered for non-collinear calculations with SOC. A penalty weight parameter $\lambda$ was set at 10 for constrained moment calculations.  The tetrahedron method with Bl\"{o}chl corrections~\cite{Blochl_1994} was employed for the BZ integration. $\Gamma$-point force constant matrix, Born effective charges and the dielectric tensor were evaluated using density functional perturbation theory (DFPT)~\cite{Baroni_2001} and PHONOPY was used for post-processing~\cite{Togo_2015}. 
We employed the Berry phase method based on the modern theory of polarisation~\cite{Vanderbilt1,Vanderbilt2} to compute the spontaneous polarisation of the experimentally observed polar magnetic structure by considering the non-polar aristotype structure as a reference.

We used the web-based ISOTROPY software suit which applies group-theoretical methods in describing distortions and possible phase transitions in crystalline materials~\cite{isotropy}. Magnetic space groups were determined from the output of VASP calculations using FINDSYM software~\cite{findsym1,findsym2} by varying the tolerance for the magnetic moments between 0.001 to 0.1. Mode analysis was done via ISODISTORT which was also employed to visualise and explore the structural and magnetic distortions~\cite{isodistort1,isodistort2}. VESTA was used to visualise the crystal structures and spin configurations~\cite{VESTA}.

The strength of the lattice-mediated linear ME coupling was computed by displacing atoms away from their zero electric field equilibrium positions and calculating net spin moments for each of the displaced structures. The electric field ($\mathbf{E}$)-induced polar displacements were determined within the formalism introduced by \'{I}\~{n}iguez~\cite{Iniguez_2008}: $ d^i_\eta = \sum_{\xi, j} C^{-1}_{\eta \xi, ij} Z^{*}_{\xi, jk} E^k $ (in the atomic basis~\cite{Spaldin_2012,Lepetit_2016}). Here, $\eta$, $\xi$ denote the atomic labels and $i$, $j$, $k$ = $x, y, z $ represent spatial coordinates. $C^{-1}_{\eta \xi, ij}$ are the matrix elements of the inverse $\Gamma$-point force constant matrix and $Z^{*}_{\xi, jk}$ are the components of the Born effective charge (BEC) tensor of the $\xi$-th atom. 

\section{Crystal Structure and Symmetry Analysis}
Phase transitions in crystalline solids, accompanied by structural distortions and/or magnetic ordering, lead to lowering of the crystal symmetry from the high-symmetry parent structure. The resulting low-symmetry daughter phases are generally related to the aristotype phase by group-subgroup relationship. In Landau's model of phase transitions, such distortions are defined in terms of order parameters and can be described by the irreducible representations (irreps) of the parent space group~\cite{Cao_2008}. 
\begin{figure}[h]
\centering
\includegraphics[scale=0.47]{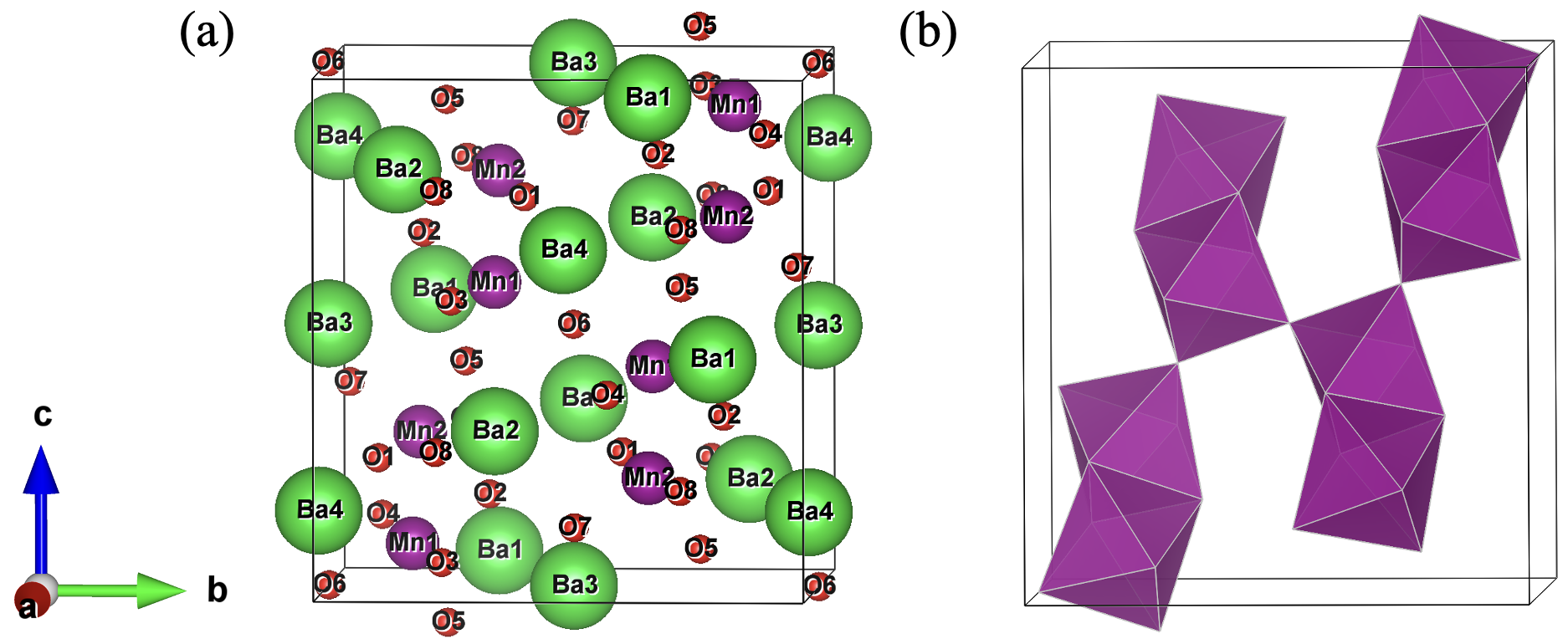}
 \caption{(a) Monoclinic crystal structure of Ba$_7$Mn$_4$O$_{15}$ with atomic site labels. Ba, Mn and O atoms are denoted by the green, purple and red spheres, respectively. As seen, Ba(3) and O(6) atoms occupy high symmetry positions. (b) Face-sharing Mn$_2$O$_9$ octahedral dimers in Ba$_7$Mn$_4$O$_{15}$ containing Mn$^{4+}$ ions.}
\label{struct}
\end{figure}

Ba$_7$Mn$_4$O$_{15}$ is isostructural to Sr$_7$Mn$_4$O$_{15}$ and crystallises in the aristotype monoclinic space group $P$2$_1/c$ (number 14) at room temperature, retaining the same symmetry down to very low temperature~\cite{Clarke_2022}. The unit cell contains two formula units with 52 atoms (hence 8 Mn atoms per unit cell) and consists of face-sharing Mn$_2$O$_9$ octahedral dimers which form strings in the $c$-plane by sharing corners~\cite{Clarke_2022}. Fig.~\ref{struct}(a) shows the  unit cell which has a monoclinic unique axis $b$. The crystal structure is non-polar, where only the Ba(3) and O(6) atoms occupy high symmetry positions, as shown in Fig.~\ref{struct}(a). The face-sharing Mn$_2$O$_9$ dimers, depicted in Fig.\ref{struct}(b), are relatively rare in comparison to common corner-sharing systems such as perovskites and can influence the magnetic interactions between the Mn$^{4+}$ ions. 

\begin{figure}[h]
\centering
\includegraphics[scale=0.47]{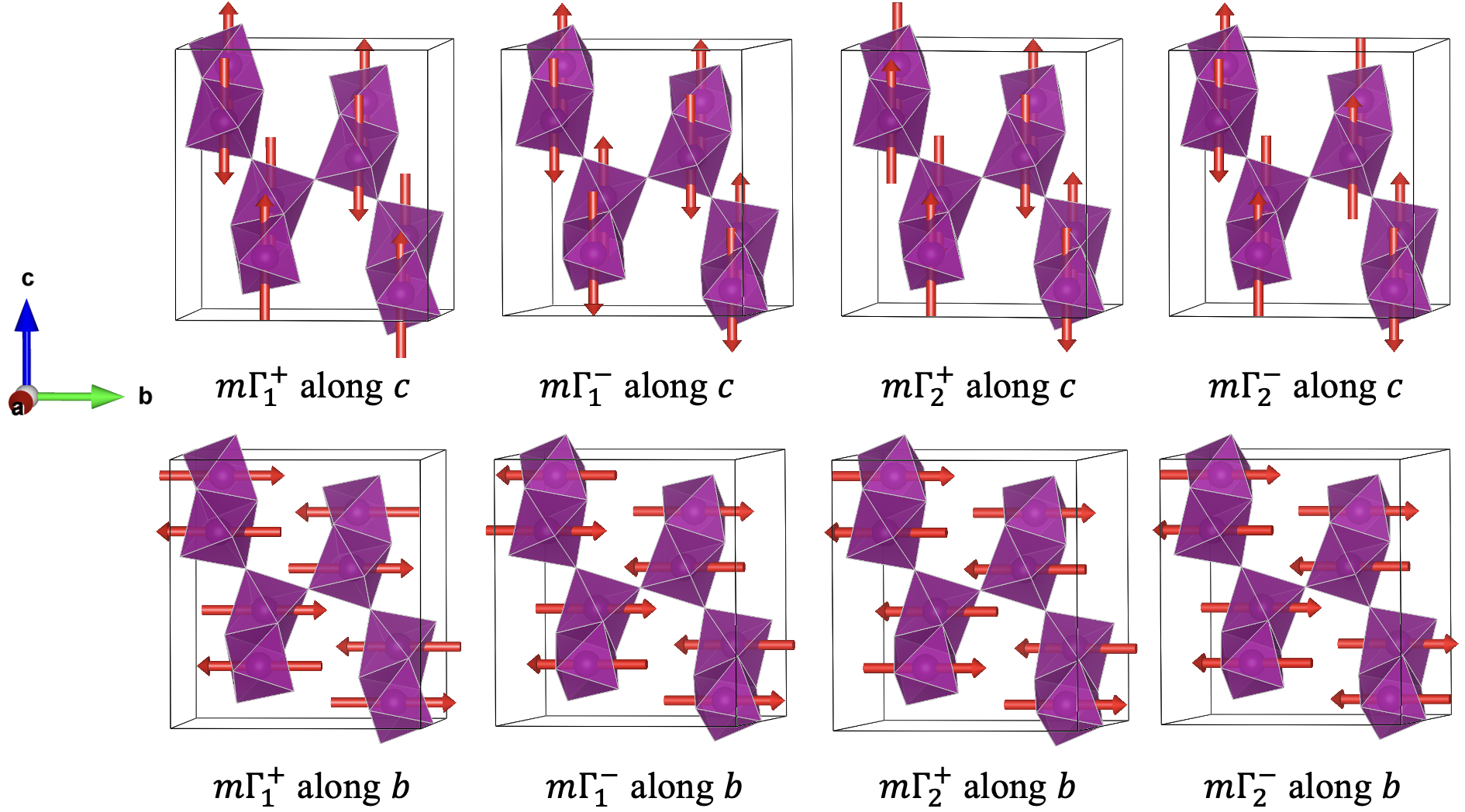} 
 \caption{Spin configurations associated with the four individual magnetic irreps present in the $P$2$_1/c$ unit cell of Ba$_7$Mn$_4$O$_{15}$. The upper and lower panels show the spin structures corresponding to each of the magnetic modes where spins are aligned along $c$ and $b$-directions, respectively. The configurations in the upper panel are degenerate within the $ac$-plane. Spin magnetic moments on the Mn$^{4+}$ ions are indicated by red arrows. The experimentally proposed $Pc$ space group results from the combination of irreps $m\Gamma^-_2$ along $b$ and $m\Gamma^+_1$ along $c$. On the other hand, combination of $m\Gamma^-_1$ along $c$ with $m\Gamma^+_2$ along $b$ leads to $Pc^\prime$ space group which has shown similar quality of fit to the NPD data in recent experiments~\cite{Clarke_2022}. There are two crystallographically unique Mn sites within the $P$2$_1/c$ unit cell and the magnetic moments across these (within the dimers) have been constrained to be antiferromagnetic in accordance with the previous experimental observations and probable strong direct exchange interactions.}
\label{irreps}
\end{figure}

Experimentally, no superstructural modes have been observed~\cite{Clarke_2022} and therefore, considering the $P2_1/c$ space group as the parent, we use ISODISTORT~\cite{isodistort1,isodistort2} to list the possible sets of magnetic irreps induced by Mn atoms. We find that in one unit cell, we have four different $\Gamma$-point magnetic irreps transforming as single magnetic propagation vectors at $\mathbf{k} = (0, 0, 0)$. Based on the fitting of the experimental neutron diffraction data, Clarke et al.~\cite{Clarke_2022} have found that for all the magnetic configurations the Mn$_2$O$_9$ dimers favour antiferromagnetically arranged Mn$^{4+}$ ions which can be explained by the strong antiferromagnetic direct exchange interaction between the half-filled $t_{2g}$ orbitals of neighbouring Mn$^{4+}$ ions. Since the individual irreps can have three Cartesian components, the spins on the Mn atoms within each irrep are allowed to move in any direction and a single mode can give rise to spin canting. The four individual magnetic modes with fixed spin alignment directions are shown in Fig.~\ref{irreps}. 

In previous experiments it is found that none of the individual magnetic modes are able to fit all the magnetic peaks observed in the NPD data and a binary combination of magnetic modes is required to accurately describe the magnetic behaviour of Ba$_7$Mn$_4$O$_{15}$~\cite{Clarke_2022}. As discussed in Ref.~\cite{Clarke_2022}, we also list the sets of magnetic space groups in Table~\ref{Table1} which result from the binary combinations of separate magnetic modes.
\begin{table}[h]
  \centering
  \caption{Magnetic space groups arising from binary combinations of magnetic modes. Order parameter direction (OPD) of each irrep is given within the parenthesis.}
  \label{Table1}
  \begin{tabular}{|c|c|c|c|c|}
    \hline
   Individual modes (OPD) &$m\Gamma^+_1 (a)$& $m\Gamma^-_1 (a)$&$m\Gamma^+_2 (a)$& $m\Gamma^-_2 (a)$\\
    \hline\rule{0pt}{1.5\normalbaselineskip}
     $m\Gamma^+_1 (a)$&$P$2$_1/c$&$P2_1$&$P\bar{1}$&$Pc$\\
     \hline\rule{0pt}{1.5\normalbaselineskip}
     $m\Gamma^-_1 (a)$&$P2_1$&$P2_1/c^\prime$&$Pc^\prime$&$P\bar{1}^\prime$\\
     \hline\rule{0pt}{1.5\normalbaselineskip}
     $m\Gamma^+_2 (a)$&$P\bar{1}$&$Pc^\prime$&$P2^\prime_1/c^\prime$&$P2^\prime_1$ \\
     \hline\rule{0pt}{1.5\normalbaselineskip}
     $m\Gamma^-_2 (a)$&$Pc$&$P\bar{1}^\prime$&$P2^\prime_1$&$P2^\prime_1/c$\\
    \hline
\end{tabular}
\end{table}
Experimentally, the best fit to the magnetic peaks comes from two different combinations: (i) $m\Gamma^-_2$ along $b$ with $m\Gamma^+_1$ along $c$ and (ii) $m\Gamma^-_1$ along $c$ with $m\Gamma^+_2$ along $b$, leading to polar space groups $Pc$ and $Pc^\prime$, respectively~\cite{Clarke_2022}. Both these combinations are found to give similar quality of fit resulting in non-collinear spin arrangements~\cite{Clarke_2022}. It should be noted that although constraining the Mn moments along $c$ has not made any difference in the experimental fits, by symmetry, the spins are free to move within the $ac$-plane~\cite{Clarke_2022}.

\section{Results and Discussion}
\subsection{Constrained moment calculations with experimental magnetic structure} 
\begin{figure}[h]
\centering
\includegraphics[scale=0.485]{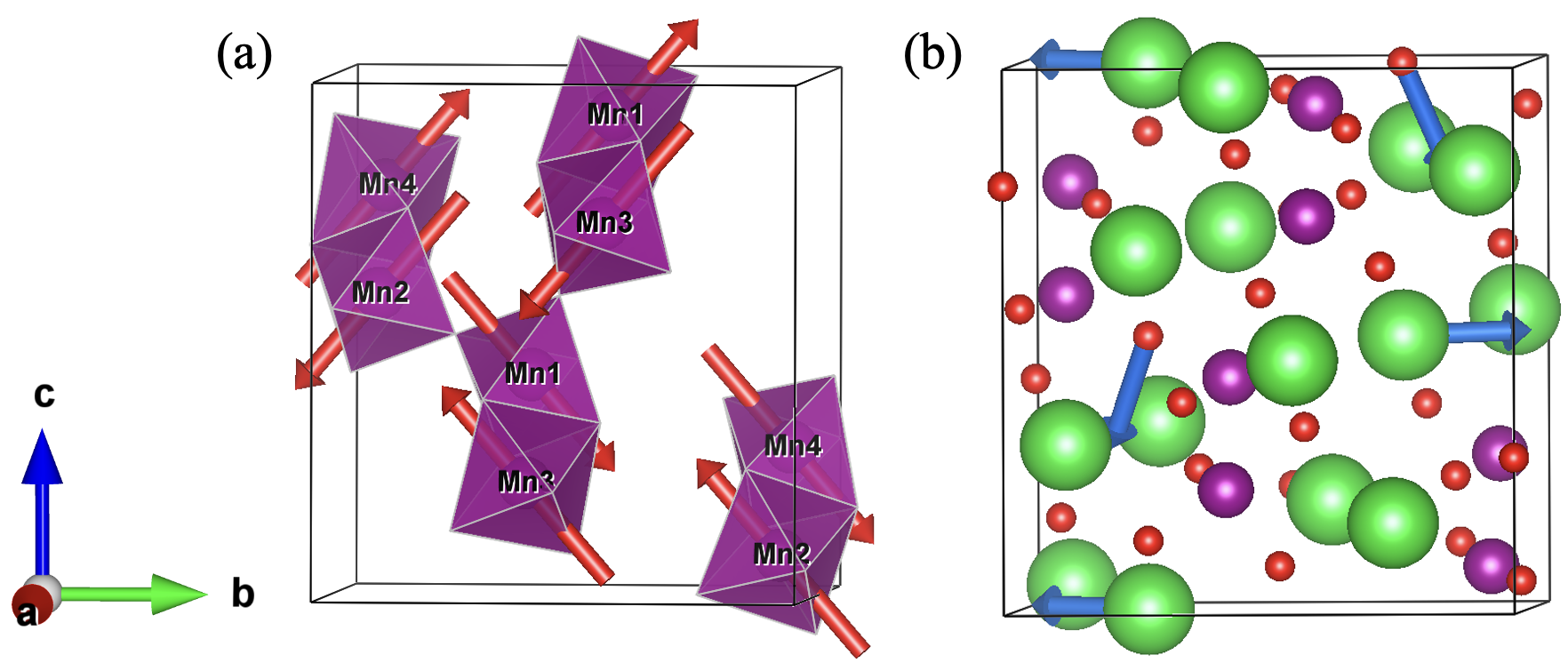}
\caption{(a) Non-collinear spin configuration in the $Pc$ space group (basis=\{(1, 0, 0), (0, 1, 0), (0, 0, 1)\}, origin shift=(0, $\frac{1}{4}$, 0) with respect to the $P$2$_1/c$ parent) observed in experiments~\cite{Clarke_2022}.  Spin magnetic moments on the Mn$^{4+}$ ions are indicated by red arrows. (b) Polar displacements (blue arrows) of the O(6) atoms (atomic label with respect to the parent structure) in the relaxed $Pc$ structure obtained from constrained moment calculations.}
\label{constrained_Pc}
\end{figure}

Earlier experiments suggest two equally probable magnetic ground states in the $Pc$ and $Pc^\prime$ space groups arising from the combinations $m\Gamma^-_2 \bigoplus m\Gamma^+_1$ and $m\Gamma^-_1 \bigoplus m\Gamma^+_2$, respectively~\cite{Clarke_2022}. In order to model the experimental magnetic structure, we perform constrained moment calculations by fixing the amplitudes of the two magnetic modes $m\Gamma^-_2$ and $m\Gamma^+_1$ according to the experimental values only for the proposed $Pc$ configuration which has been put forward as the most likely space group~\cite{Clarke_2022}. Additionally, we keep the spin directions of the modes fixed i.e. we consider the $m\Gamma^-_2$ mode along $b$ while the $m\Gamma^+_1$ mode is constrained to point along the $c$-direction~\cite{Clarke_2022}. The parity of both order parameters are odd with respect to time reversal but only one is odd with respect to inversion symmetry, thus their joint action leads to a global breaking of inversion symmetry. Furthermore, the resulting $Pc$ space group has a polar point group $m$ which is also compatible with weak ferromagnetism. The non-collinear arrangement of the spins on the four inequivalent Mn sites in the $Pc$ structure is shown in Fig.~\ref{constrained_Pc}(a).

We first fully relax the $Pc$ structure with constrained moments by setting $U$ = 2.5 eV and $J$ = 0.5 eV on the Mn$^{4+}$-$d$ orbitals which lead to lattice parameters and cell volume close to the experimental values. The lattice parameters and atomic coordinates of the fully relaxed $Pc$ structure (basis=\{(1, 0, 0), (0, 1, 0), (0, 0, 1)\}, origin shift=(0, $\frac{1}{4}$, 0) with respect to the $P$2$_1/c$ parent structure) are listed in Table~S1 of the Supplementary Information (SI)~\cite{supp}. Moreover, calculation of the total density of states reveals that the resulting $Pc$ structure is insulating and possesses a band gap of $\sim$ 0.94 eV, as shown in Fig.~S1~\cite{supp}.

Interestingly, in the fully relaxed structure, we observe polar displacements of the Ba(3) and O(6) atoms (see Fig.~\ref{constrained_Pc}(b)) which are located at high symmetry positions in the $P$2$_1/c$ phase. These polar displacements have $\Gamma^-_2$ character and are of the order of thousandths of an angstrom which could not be detected in earlier experiments~\cite{Clarke_2022}. Since the polar mode is induced by magnetic ordering, Ba$_7$Mn$_4$O$_{15}$ is classified as a type-II multiferroic. 
We further calculate the ferroelectric polarisation of the DFT-relaxed $Pc$ phase using the Berry phase method based on the modern theory of polarisation which shows that the $Pc$ phase of Ba$_7$Mn$_4$O$_{15}$ possesses a spontaneous polarisation of $\mathbf{P} = (177.24,  0.00,  251.30)$ $\mu$C/m$^2$. The polarisation components are sizeable and comparable to known type-II multiferroics~\cite{Kimura_2003,Khomskii_2009}. Therefore, experimentally it should be possible to measure the spontaneous polarisation and ferroelectric switching in Ba$_7$Mn$_4$O$_{15}$ provided the sample is insulating enough at low temperatures. INVARIANTS~\cite{invariants1,invariants2} analysis reveals that in the $Pc$ space group the $\Gamma^-_2$ polar mode couples to the magnetic modes linearly giving rise to a trilinear coupling term of the form $m\Gamma^-_2 \bigoplus m\Gamma^+_1 \bigoplus \Gamma^-_2$ in the free energy expansion.  

Although we find that constraining the magnetic moments according to the experimental magnetic configuration in the $Pc$ space group indeed leads to a multiferroic phase in Ba$_7$Mn$_4$O$_{15}$, from symmetry considerations, there is no reason for the magnetic moments to be constrained along a fixed direction. Moreover, another phase in the $Pc^\prime$ space group arising from the combination of magnetic irreps $m\Gamma^-_1$ along $c$ and $m\Gamma^+_2$ along $b$ has been found to show similar quality of fit to the NPD data in previous experiments~\cite{Clarke_2022}. Therefore, in order to find the minimum energy magnetic configuration from first-principles, we perform a series of DFT calculations starting from several different non-collinear and collinear spin configurations as described in the next subsection.   

\subsection{Magnetic ground state from self-consistent first-principles calculations}
Considering the experimental low temperature (100 K) nuclear structure in the $P$2$_1/c$ space group~\cite{Clarke_2022}, we perform full relaxation of the unit cell in the collinear $m\Gamma^{+}_1$ configuration for a number of $U$ and $J$ parameters varied within reasonable range and find that $U$ = 2.5 eV and $J$ = 0.5 eV lead to experimentally comparable cell parameters. Relaxation with this collinear spin configuration retains the high-symmetry $P$2$_1/c$ space group of the unit cell. The fully relaxed cell parameters and atomic coordinates, listed in Table~S2~\cite{supp}, match well with the values in the experimental $P$2$_1/c$ structure~\cite{Clarke_2022} and there is no deviation of the Ba(3) and O(6) atoms from their high symmetry positions. With these values of $U$ and $J$, we obtain magnetic moment magnitudes of $\sim$ 2.79 $\mu_B$ and $\sim$ 2.77 $\mu_B$ of the Mn1 and Mn2 atoms, respectively, which are slightly larger than the experimentally observed values of $\sim$ 2.3 $-$ 2.4 $\mu_B$~\cite{Clarke_2022}. The fully relaxed  structure in the $m\Gamma^{+}_1$ configuration remains insulating with an energy band gap of $\sim$ 1.38 eV, as shown in Fig.~S2~\cite{supp}.
\begin{figure}[h]
\centering
\includegraphics[scale=0.79]{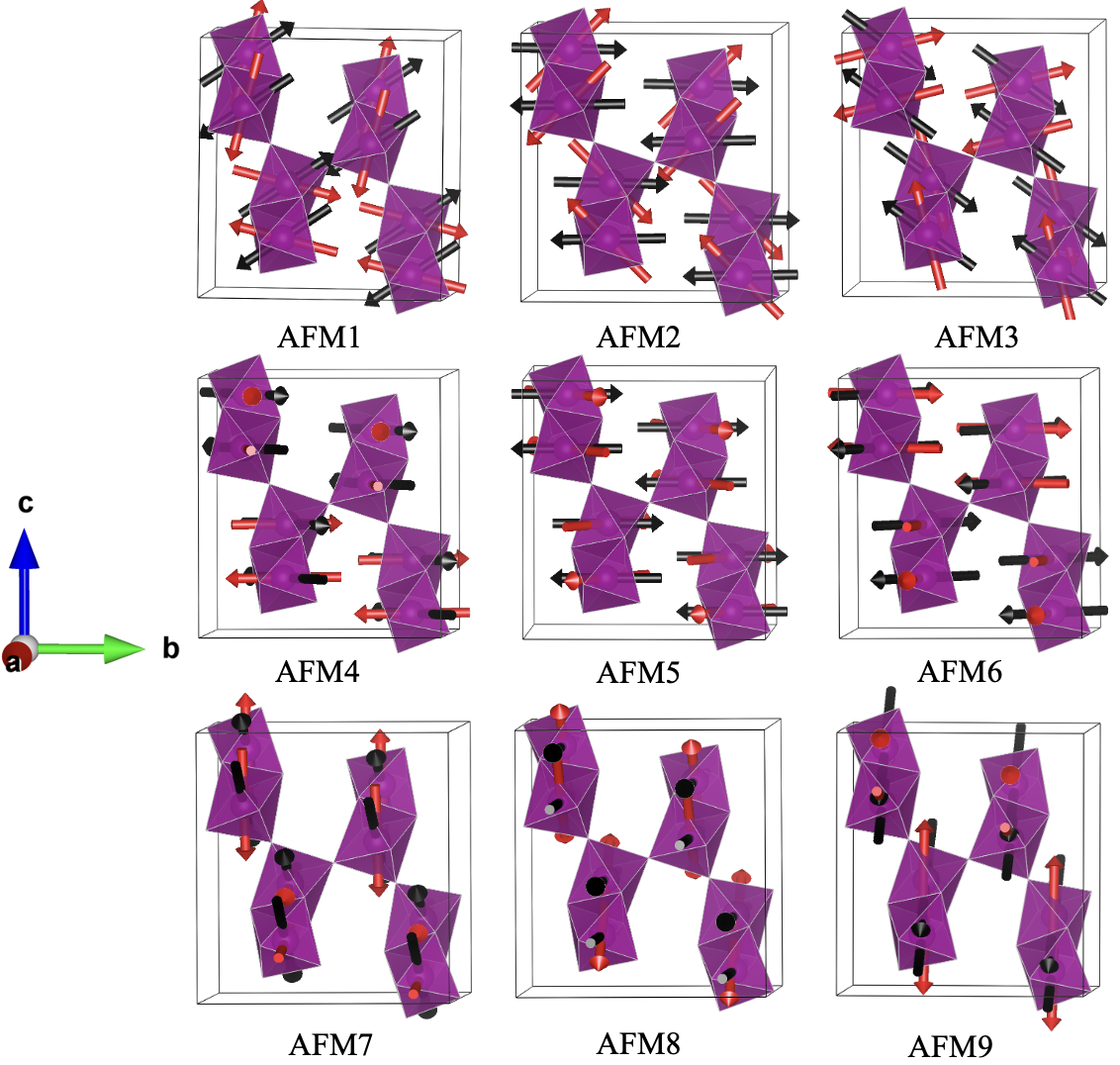} 
\caption{Initial and final spin configurations in our non-collinear DFT calculations, where initial spin moments, indicated by red arrows, lie within the $bc$- (upper panel), $ab$- (middle panel) and $ac$-planes (lower panel). AFM2 (initial) corresponds to the experimentally suggested $Pc$ configuration. Final spin magnetic moments on the Mn$^{4+}$ ions are indicated by black arrows. As seen, in all cases, self-consistent relaxation of the spin densities leads to (nearly) collinear final spin arrangements.}
\label{Pc-nc}
\end{figure} 

\subsubsection{Non-collinear magnetic calculations}
We next consider three different initial non-collinear antiferromagnetic (AFM) spin configurations, where the magnetic moments are fixed to lie within the $bc$-plane, as shown in Fig.~\ref{Pc-nc}. One of them being the experimentally suggested $Pc$ structure (AFM2) with only $m\Gamma^-_2$ and $m\Gamma^+_1$ modes, whereas in the other two structures (AFM1 and AFM3), all the four magnetic modes $m\Gamma^-_2$,  $m\Gamma^+_1$, $m\Gamma^-_1$ and $m\Gamma^+_2$ with different amplitudes are included. In this case, instead of constraining the moment amplitudes and directions, we let the spins relax self-consistently including SOC. Within the dimers, antiparallel arrangement of spins are considered because of the strong antiferromagnetic direct exchange coupling between neighbouring Mn$^{4+}$ ions, as also observed in experiments~\cite{Clarke_2022}. In order to explore all possible magnetic configurations with different amplitudes and OPDs of the four individual magnetic modes, we further consider non-collinear spin structures where the initial spin moments lie within the $ab$ (AFM4$-$AFM6) and the $ac$-planes (AFM7$-$AFM9) separately. Fig.~\ref{Pc-nc} illustrates the nine different initial and final antiferromagnetic configurations. We relax the spins self-consistently starting from each of the initial configurations and determine the magnetic space groups of the resulting structures from the final magnetic moments using FINDSYM~\cite{findsym1,findsym2}. Canting angles of the spins within the individual Mn$_2$O$_9$ dimers are also calculated. Results of our self-consistent spin relaxation calculations for these non-collinear spin structures are summarised in Table~S3. 

\begin{figure}[h]
\centering
\includegraphics[scale=0.44]{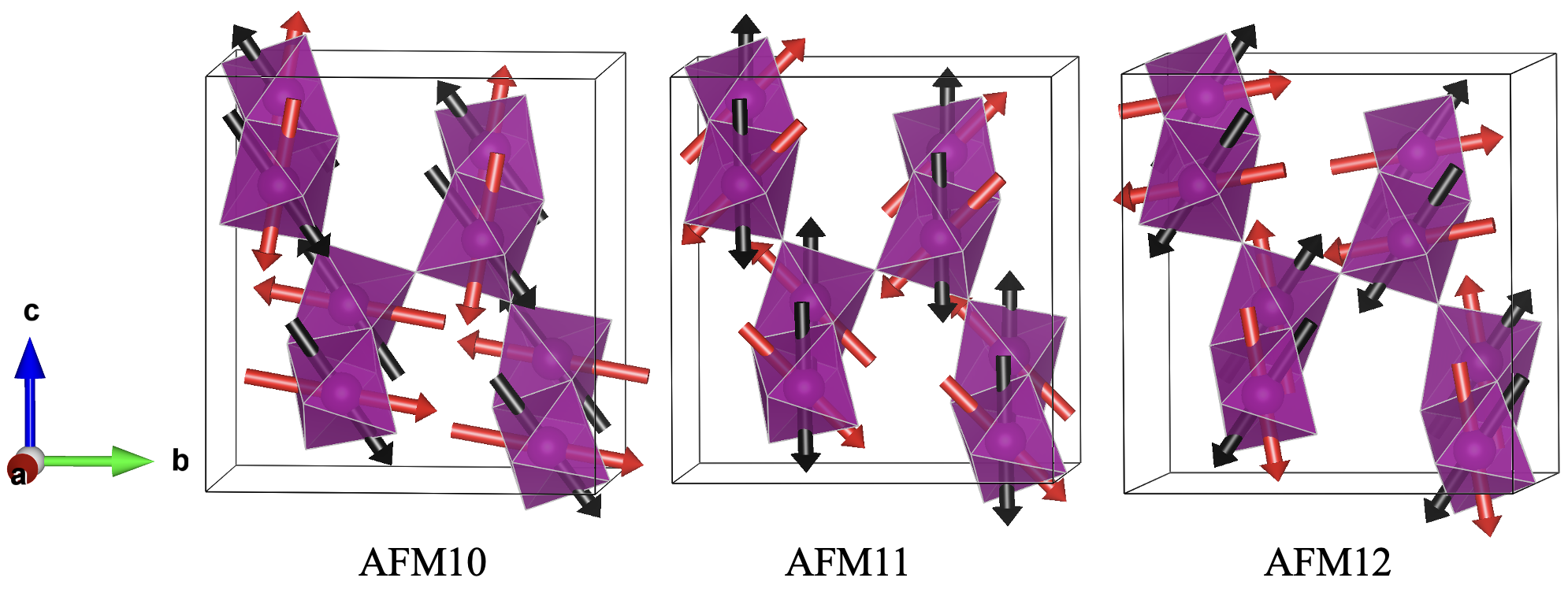}
\caption{Initial and final non-collinear spin configurations in our DFT calculations where the initial Mn spins, indicated by red arrows,  are allowed to lie within the $bc$ plane. Here, the dimers initially have different relative spin orientations compared to those in Fig.~\ref{Pc-nc}. AFM11 (initial) corresponds to the experimentally proposed $Pc^\prime$ configuration. Final spin magnetic moments on the Mn$^{4+}$ ions are indicated by black arrows. As in Fig.~\ref{Pc-nc}, all the  non-collinear calculations finally lead to (nearly) collinear spin configurations.}
\label{Pc'-nc}
\end{figure}

Since Clarke et al. have also suggested another possible magnetic structure in the $Pc^\prime$ space group arising from the combination of $m\Gamma^-_1$ irrep along $c$ and $m\Gamma^+_2$ irrep along $b$~\cite{Clarke_2022}, we again consider three different initial non-collinear spin structures within the $bc$-plane as before, however, with different relative spin orientations between the dimers. One of the initial configurations is considered in the $Pc^\prime$ space group (AFM11) with $m\Gamma^-_1$ and $m\Gamma^+_2$ irreps, whereas, in the other two initial configurations (AFM10 and AFM12) all the four individual magnetic irreps are included with different proportions (see Fig.~\ref{Pc'-nc}). 

It is found that even if we start with a non-collinear spin structure, the spins spontaneously relax into a (nearly) collinear configuration in each case, as shown in Figs.~\ref{Pc-nc} and \ref{Pc'-nc}, see Tables~S3 and S4~\cite{supp} for details. Investigation into the directions of the final magnetic moments of the Mn atoms reveal that in almost all cases, the Mn atoms within the dimers are slightly canted. Among the 12 AFM configurations, the two lowest energy final configurations (AFM2 and AFM5) correspond to spin arrangements where magnetic moments have a large component along the $b$-axis and very small moments along the other two directions. Both these configurations lead to a magnetic space group $P\bar{1}^\prime$, which is different from the experimentally proposed magnetic structures.  

Since all the non-collinear calculations finally lead to (nearly) collinear spin configurations, in the next step, we investigate a number of different initial collinear configurations as outlined in the next subsection.
\subsubsection{Collinear magnetic calculations}
\begin{figure}[h]
\centering
\includegraphics[scale=0.49]{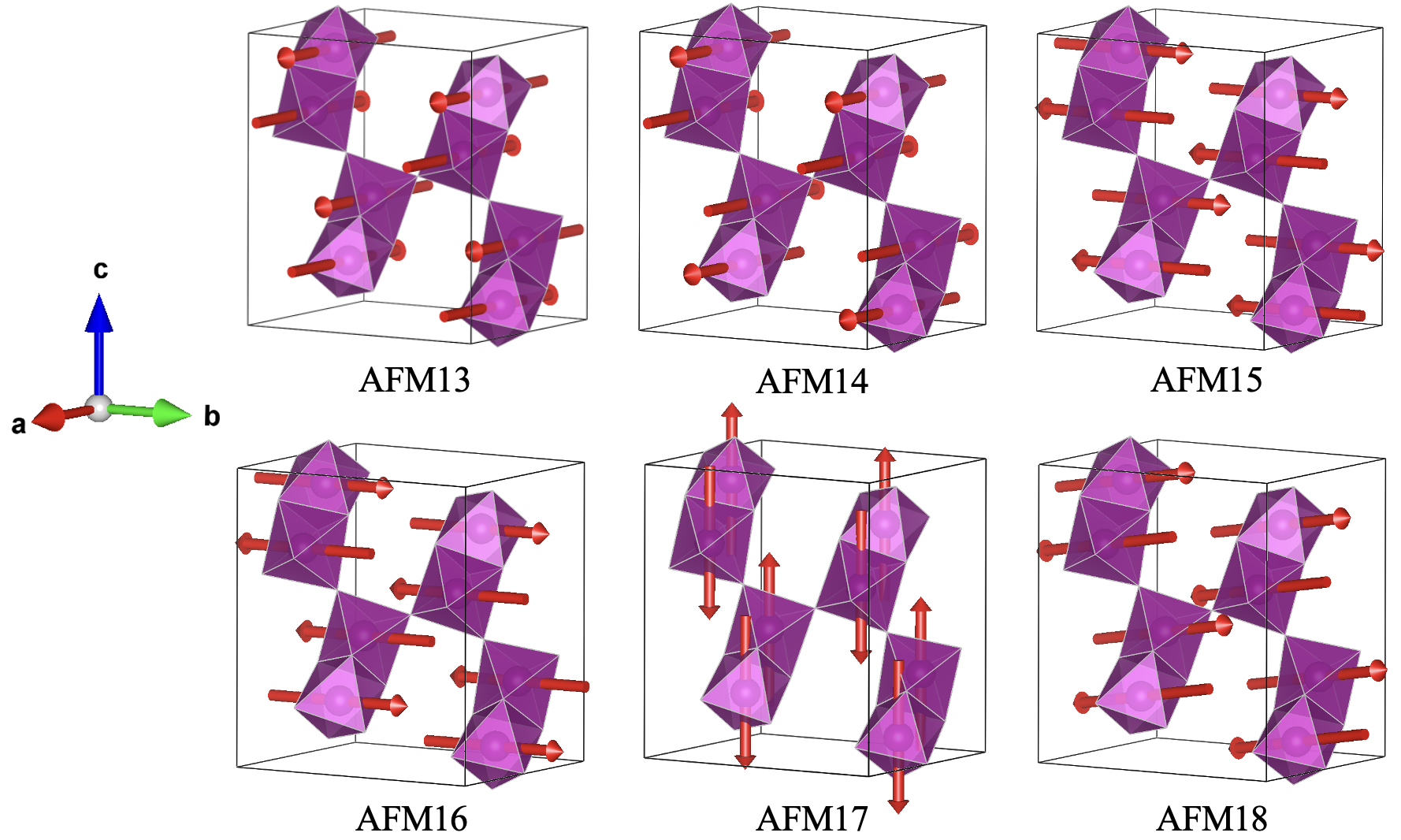} 
 \caption{Initial collinear spin configurations with the Mn spins pointing along certain crystallographic directions. Initial spin magnetic moments on the Mn$^{4+}$ ions are indicated by red arrows. AFM18 corresponds to a spin structure where the spins are aligned along a general diagonal direction with a larger component along $b$. After self-consistent spin relaxation the spin moments remain collinear with small finite spin canting angles (not shown), see SI~\cite{supp} for details.}
\label{coll}
\end{figure}

The lowest energy configurations obtained from the non-collinear calculations show that the magnetic moments tend to point towards the $b$-direction, however, very small moments also arise in the other two directions leading to canted Mn spins within the dimers. Therefore, in order to check the stability of the spin canting, six initial collinear configurations are considered, as shown in Fig.~\ref{coll}. For all the calculations, we include SOC effect and allow the spins to relax self-consistently as before. 

Investigation into the total energies of the final spin-relaxed collinear structures reveal that in the lowest energy configuration, the magnetic moments point along a general diagonal direction with a large component along the $b$-direction. The final spin moments, total energies and spin canting angles are given in Table~S5~\cite{supp}. Interestingly, we notice that even when we start with a configuration with spins directed solely along $a$, $b$ and $c$ axes, spin moments spontaneously arise in the other two directions, indicating that the spin cantings are real and appear in almost all cases. The ground state magnetic structure is found to have $P{\bar{1}}^\prime$ space group (basis = \{(1, 0, 0), (0, 1, 0), (0, 0, 1)\}, origin shift = (0, $\frac{1}{2}$, $\frac{1}{2}$) with respect to the aristotype), in contrast to the experimentally suggested $Pc$ or $Pc^\prime$ space group. In the final ground state, the Mn spins inside the dimers are slightly canted with a canting angle of $\sim 0.10\degree$ (see SI for details~\cite{supp}). Mode analysis of the $P{\bar{1}}^\prime$ magnetic structure performed with ISODISTORT~\cite{isodistort1,isodistort2} shows the presence of a binary combination of modes, where the dominant contribution ($\sim$ 81\%) arises from the $m\Gamma^-_2$ mode along $b$. The other mode with $\sim$ 19\% contribution is found to possess $m\Gamma^-_1$ character along $c$ and as a result, the final ground state remains non-polar.
\subsubsection{Effects of $U$ and $J$ on the magnetic ground state}
In order to ascertain the stability of the magnetic ground state obtained from our self-consistent DFT calculations, we repeat calculations with three different sets of $U$ and $J$ values for some of the final spin configurations which are very close in energy, including the one with initial $Pc$ configuration. We consider three extreme sets of $U$ and $J$ values: (i) $U = 0.0$ eV, $J = 0.0$ eV; (ii) $U = 5.0$ eV, $J = 1.0$ eV; (iii) $U = 8.0$ eV, $J = 1.0$ eV. Calculations of the total density of states show that for all three sets of $U$ and $J$ values, the structure remains insulating without a qualitative change in the overall density of states. The details of the ground state spin configurations for different $U$ and $J$ values are listed in Table~S6~\cite{supp}. We find that in all three cases, the spins spontaneously relax into a $P{\bar{1}}^\prime$ magnetic space group as found from calculations with $U = 2.5$ eV and $J = 0.5$ eV. However, the amplitudes of the magnetic modes and spin canting angles are larger for higher $U$ and $J$, as shown in Table~\ref{tab-UJ}. 
Thus, we find that the ground state magnetic configuration in the $P{\bar{1}}^\prime$ space group with a larger spin component along $b$ is stable with respect to the $U$ and $J$ parameters. The ground state spin structure from self-consistent DFT calculations is shown in Fig.~\ref{g.s.}.\\

\begin{minipage}{\textwidth}
  \begin{minipage}[b]{0.45\textwidth}
    \centering
    \includegraphics[scale=0.475]{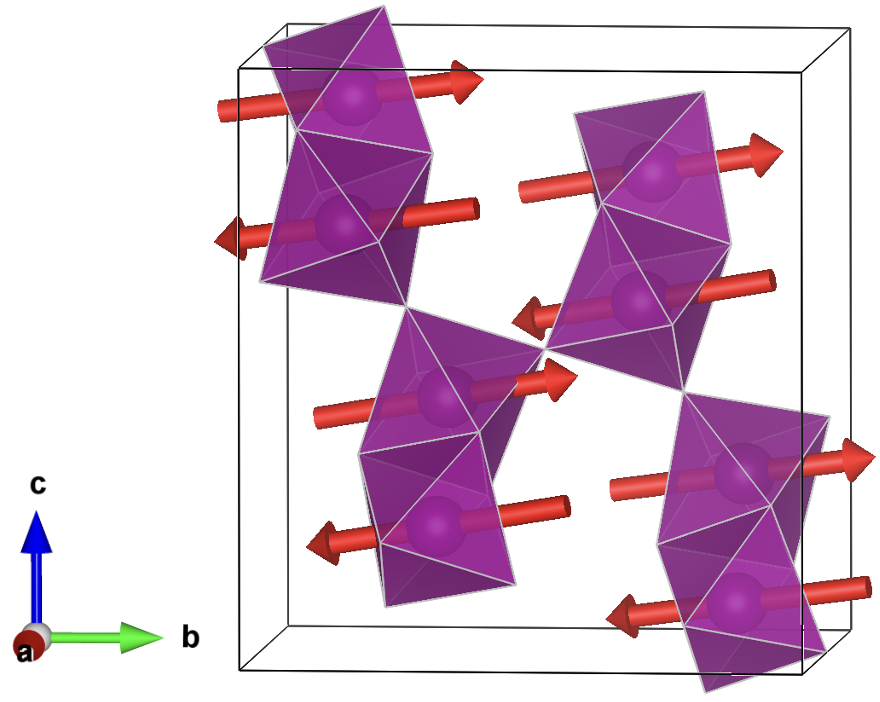}
    \captionof{figure}{Ground state magnetic structure in the $P{\bar{1}}^\prime$ space group (basis = \{(1, 0, 0), (0, 1, 0), (0, 0, 1)\}, origin shift = (0, $\frac{1}{2}$, $\frac{1}{2}$) with respect to the $P$2$_1/c$ parent structure) obtained from self-consistent DFT calculations. Spin magnetic moments on the Mn$^{4+}$ ions are indicated by red arrows.}
    \label{g.s.}
  \end{minipage}
  \hspace{0.56cm}
  \begin{minipage}[b]{0.45\textwidth}
    \centering
    \begin{tabular}{|c|c|c|}\hline
      $U$ and $J$  & Mode Amplitude & $\theta$ \\
      (eV) & ($\mu_B$) &  ($\degree$)\\\hline
      $U$ = 0.0 &$m\Gamma^-_1$ = 0.002&0.08\\
        $J$ = 0.0 &$m\Gamma^-_2$ = 7.215 &  \\\hline
        $U$ = 2.5 &$m\Gamma^-_1$ = 1.757&0.10\\
        $J$ = 0.5 &$m\Gamma^-_2$ = 7.601& \\\hline
        $U$ = 5.0 &$m\Gamma^-_1$ = 1.883&0.14\\
        $J$ = 1.0 &$m\Gamma^-_2$ = 8.153 &  \\\hline
        $U$ = 8.0 &$m\Gamma^-_1$ = 2.101&0.13\\
        $J$ = 1.0 &$m\Gamma^-_2$ = 9.115&  \\   \hline           
      \end{tabular}
      \captionof{table}{Amplitudes of the magnetic modes present in the $P{\bar{1}}^\prime$ ground state configuration obtained from self-consistent DFT calculations. Mode amplitudes are calculated with respect to the relaxed $P2_1/c$ parent structure for different $U$ and $J$. $\theta$ is the average canting angle of the Mn moments within the dimers.}
      \label{tab-UJ}
    \end{minipage}
  \end{minipage}\\\\

It is important to note that we have checked the $P\bar{1}^\prime$ model obtained from self-consistent DFT calculations against the original experimental NPD data in Ref.~\cite{Clarke_2022}, however, no satisfactory fit has been obtained. Therefore, the discrepancies between the experimental and DFT-calculated magnetic ground states do not necessarily appear due to the wrong assignment of the experimental magnetic structure. The most likely origin of this discrepancy is the hidden chemical complexity in the experimental nuclear structure arising from disorder or anion non-stoichiometry. Oxygen vacancy defects are quite common in oxides and could play a significant role in determining the electronic and magnetic properties of oxide systems~\cite{Altmeyer_2016,Dey_2018}. Sr$_7$Mn$_4$O$_{15}$ and the related cation-substituted phases have been shown to be susceptible to oxygen vacancies~\cite{Hayward_2007}. Presence of oxygen vacancies  can lead to a change in the oxidation state of the Mn$^{4+}$ ions, thereby affecting the magnetic exchange interactions between the Mn ions which can in turn lead to a change in the second magnetic mode with smaller contribution. On the other hand, although we have explored the effect of different sets of Hubbard $U$ and exchange $J$ parameters on the magnetic ground state within the PBEsol+$U$ framework, inclusion of more accurate Hybrid functionals~\cite{Hybrid} or use of more sophisticated dynamical mean-field theory (DMFT) methods~\cite{DMFT} might come closer to resolving the true magnetic ground state of Ba$_7$Mn$_4$O$_{15}$. This is beyond the scope of the present work, but would be encouraged for future study.
\subsection{Linear magnetoelectric coupling in Ba$_7$Mn$_4$O$_{15}$}
The linear ME effect refers to the induction of (or change in) magnetisation $M_j$ (polarisation $P_i$) by an external electric field $E_i$ (magnetic field $H_j$) according to the linear relation: $\mu_0 M_j = \alpha_{ij}E_i$ ($P_i = \alpha_{ij}H_j$), where $\mu_0$ is the vacuum permeability and $i, j$ are spatial coordinates. The linear ME tensor $\bm{\alpha}$ consists of different microscopic contributions coming from the ionic, electronic and strain-mediated couplings to the spin and orbital moments of electrons~\cite{Birol_2012}. In our work, only the lattice-mediated spin contribution to linear ME tensor was computed as it often plays the  dominant role in determining the magnitude of the total ME response~\cite{Cheong_2007,Iniguez_2008,Iniguez_2009}.

Calculation of the $\Gamma$-point phonons in the $m\Gamma^+_1$ magnetic configuration without SOC shows the absence of any imaginary phonon frequencies. Moreover, it is found that there are 41  $A_u$ ($\Gamma^-_1$) phonon modes and 40 $B_u$ ($\Gamma^-_2$ ) phonon modes (per unit cell excluding the acoustic modes) at the zone centre which are infrared (IR) active and can couple to the external electric field:
\begin{equation}
\Gamma_{\text{IR}} = 41 A_u + 40 B_u.
\end{equation}

$\mathbf{E}$-induced polar displacements are obtained from the $\Gamma$-point force constant matrix and Born effective charges (BEC) calculated in the $m\Gamma^+_1$ magnetic configuration. Collinear spin ordering along $c$ is considered in absence of SOC to lower the computational cost. However, inclusion of SOC and use of other magnetic configurations are found to yield almost identical values of the force constant matrix elements and BEC. The dielectric permittivity tensor and BEC computed with the $m\Gamma^+_1$ spin arrangements are given in Tables S7 and S8~\cite{supp}. 
\begin{figure}[h]
\centering
\includegraphics[scale=0.485]{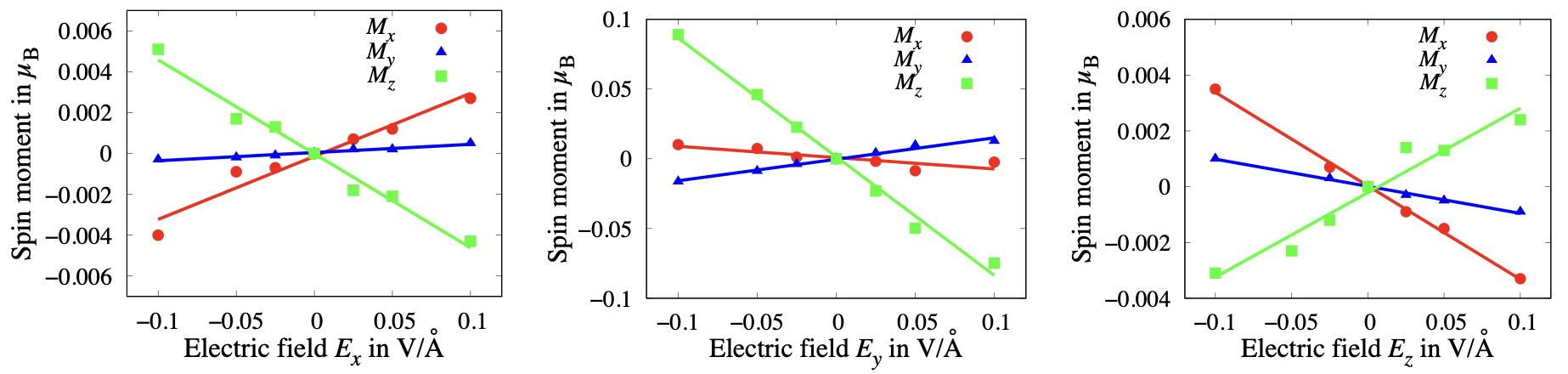}
 \caption{Calculated net spin moment per unit cell as a function of external electric field applied along the three Cartesian directions in the $P\bar{1}^\prime$ ground state of Ba$_7$Mn$_4$O$_{15}$. Lines represent the linear fits to the data. }
\label{MEC}
\end{figure}

We calculate the strength of the linear ME coupling for the non-polar $P\bar{1}^\prime$ ground state including SOC for $U = 2.5 $ eV and $J = 0.5$ eV. Due to the triclinic symmetry of the ground state magnetic structure, all the nine non-vanishing components of the linear ME tensor are independent. We, therefore, apply electric fields along the three Cartesian directions and compute the net canted spin moments arising because of the SOC effect. As shown in Fig.~\ref{MEC}, the calculated net spin moments per unit cell exhibit a linear trend with the applied electric fields demonstrating the linear ME behaviour of Ba$_7$Mn$_4$O$_{15}$. Note that in the $P\bar{1}^\prime$ structure, application of an external electric field induces a weak ferromagnetic component (along the $b$ direction) associated with the $m\Gamma^+_1$ irrep which is present in the experimentally proposed $Pc$ space group.   
\begin{table}[h!]
  \centering
  \caption{Nine independent non-vanishing components of the linear ME tensor $\alpha_{ij} ~(i, j = x, y, z)$ calculated for the non-polar $P\bar{1}^\prime$ ground state of Ba$_7$Mn$_4$O$_{15}$. Components of $\bm{\alpha}$ and their uncertainties from the linear fit are given in ps/m and Gaussian units (g.u.).}
  \label{tab-MEC}
  \begin{tabular}{|c c c|c c c|}
    \hline
    \hline
    $\alpha_{ij}$&ps/m &$10^{-4}$ g.u.& $\alpha_{ij}$&ps/m &$10^{-4}$ g.u. \\
    \hline
    $\alpha_{xx}$&0.047 $\pm$ 0.005&0.14 $\pm$ 
 0.02&$\alpha_{xy}$&0.006 $\pm$ 0.001&0.019 $\pm$ 0.002\\
    
    $\alpha_{xz}$&-0.070 $\pm$ 0.005 &-0.21 $\pm$ 0.01& $\alpha_{yx}$&-0.12 $\pm$ 0.03&-0.4 $\pm$ 0.1\\
    $\alpha_{yy}$&0.24 $\pm$ 0.01&0.71 $\pm$ 0.04&$\alpha_{yz}$&-1.30 $\pm$ 0.06&-3.9 $\pm$ 0.2\\
    $\alpha_{zx}$&-0.051 $\pm$ 0.001&-0.154 $\pm$ 0.004&$\alpha_{zy}$&-0.015 $\pm$ 0.001&-0.044 $\pm$ 0.002\\
    $\alpha_{zz}$&0.046 $\pm$ 0.005&0.14 $\pm$ 0.02&&&\\
    \hline
\end{tabular}
\end{table}

Using a linear fit to the individual responses and computing the gradient, we calculate the linear ME tensor of Ba$_7$Mn$_4$O$_{15}$ in the non-polar $P\bar{1}^\prime$ magnetic ground state. All the independent components of $\bm{\alpha}$ are shown in Table~\ref{tab-MEC}. It is interesting to note that the largest component of the linear ME tensor has a sizeable magnitude ($\sim$ 3.90 $\times 10^{-4}$ Gaussian units) which is three times the value predicted for the classic magnetoelectric Cr$_2$O$_3$ ($\sim$ 1.30 $\times 10^{-4}$ Gaussian units)~\cite{Iniguez_2008} and comparable to that of the well-known multiferroic BiFeO$_3$ ($\sim$ 5 $\times 10^{-4}$ Gaussian units)~\cite{Iniguez_2009}. 
 
\section{Summary and Conclusions}
In summary, using first-principles DFT calculations we have studied the ground state magnetic behaviour of Ba$_7$Mn$_4$O$_{15}$ which has been classified as a type-II multiferroic in earlier experiments. Our constrained moment calculations with the proposed experimental magnetic configuration in the $Pc$ space group shows the spontaneous emergence of a polar mode with significant ferroelectric polarisation comparable to other known type-II multiferroics. Moreover, we are able to identity the character of the polar displacements induced by the experimental magnetic ordering. However, self-consistent relaxation of the spin densities without any external constraints leads to a non-polar magnetic ground state in the $P\bar{1}^\prime$ space group where two magnetic modes transforming as distinct irreps coexist. The dominant magnetic mode $m\Gamma^-_2$ along $b$ is consistent with one of two previously proposed magnetic models from experiments~\cite{Clarke_2022}. However, in the $P\bar{1}^\prime$ ground state, the second magnetic mode transforms as the $m\Gamma^-_1$ irrep (along $c$) which is different from the $m\Gamma^+_1$ irrep (along $c$) found in previous experiments~\cite{Clarke_2022}. Since the combination $m\Gamma^-_2 \bigoplus m\Gamma^-_1$ does not break spatial inversion, our self-consistent calculations show the absence of a type-II multiferroic ground state. The origin of the disagreement between experiment and simulation is unclear at present but possibly points to additional complexity in the nuclear structure such as anion vacancies and disorder. Nonetheless, the $P\bar{1}^\prime$ ground state is found to exhibit a strong linear ME coupling comparable to the well-known multiferroic BiFeO$_3$~\cite{Iniguez_2009}, suggesting strategies to design new linear
magnetoelectrics.   

\section*{Acknowledgements}
U.D. and N.C.B. acknowledge the Leverhulme Trust for a research project grant (Grant No. RPG-2020-206). This work used the ARCHER2 UK National Supercomputing Service (https://www.archer2.ac.uk) and the Hamilton HPC Service of Durham University. M.S.S. acknowledges the Royal Society for a fellowship (UF160265) and EPSRC grant ``Novel Multiferroic Perovskites through Systematic Design" (EP/S027106/1) for funding.

\section*{References}
\bibliography{bmo}
\clearpage

\section*{\centering{\Large{Supplementary Information}}}
\begin{table}[h]
  \centering
  \caption*{Table S1. Lattice parameters and atomic coordinates of the fully relaxed $Pc$ structure (basis=\{(1, 0, 0), (0, 1, 0), (0, 0, 1)\}, origin=(0, $\frac{1}{4}$, 0) with respect to the $P$2$_1/c$ parent structure) with magnetic moments constrained to the experimental values and directions~\cite{Clarke_2022}. Constrained moment relaxation calculations are performed including spin-orbit coupling (SOC) and using $U$ = 2.5 eV, $J$ = 0.5 eV for the Mn-$d$ orbitals.}
  \begin{tabular}{|c|c|c|c|c|c|}
    \hline
    Cell parameters& Fully relaxed values& Atomic sites& x &y &z\\
    \hline\rule{0pt}{1.0\normalbaselineskip}
     &&Ba1&-0.00256&0.43409& -0.03406\\
     &&Ba2&0.00259 & -0.06595 &0.53406\\
     &&Ba3&0.34516 & 0.40147& 0.69537 \\
     $a$ (\AA)&7.15653&Ba4&0.65482 & -0.09856& 0.80465\\
     &&Ba5&0.49993 & 0.75010&  0.49998 \\
     $b$ (\AA)& 10.01215&Ba6&0.17837 & 0.25010&  0.33663\\
     &&Ba7&0.82174 & 0.75016 & 0.16344\\
     $c$ (\AA)& 10.66583&Mn1&0.57478 & 0.41551 & 0.42554\\
     &&Mn2&0.42528 & -0.08458& 0.07451\\
     $\alpha$ ($\degree$)& 90.00000&Mn3&0.76928 & 0.41976 & 0.22384\\
     &&Mn4&0.23067 & -0.08022 &0.27612\\
     $\beta$ ($\degree$) & 92.05227&O1&0.51330 & 0.35011 & 0.25773\\
     &&O2&0.48669 & 0.85005 & 0.24223\\
     $\gamma$ ($\degree$)& 90.00000&O3&0.67126 & 0.43332 & 0.83619\\
     &&O4&0.32875 & -0.06667 &0.66387\\
     $V_{\rm{cell}}$ ({\AA}$^3$) & 763.74054&O5&0.34275 & 0.49008 & 0.44642\\
     &&O6&0.65737 & -0.01004 &0.05367\\
     &&O7&0.82552 & 0.33825 & 0.38615\\
     &&O8&0.17447  &0.83825 & 0.11381\\
     &&O9&0.67263 & 0.48176 & 0.58171\\
     &&O10&0.32750 & -0.01794 &-0.08163\\
     &&O11&0.49988 & 0.75005 & -0.00012\\
     &&O12&0.83662 & 0.27227 & 0.13304\\
     &&O13&0.16317  &0.77224  &0.36684\\
     &&O14&0.00207 & 0.49939 & 0.21387\\
     &&O15&-0.00203 &-0.00052 &0.28614\\
    \hline
\end{tabular}
\end{table}
\begin{figure}[h]
\centering
\includegraphics[scale=1.3]{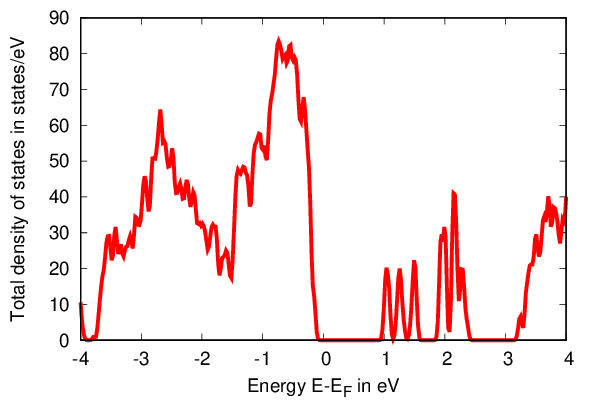} 
\caption*{Figure S1. Total density of states calculated with $U$ = 2.5 eV and $J$ = 0.5 eV for the relaxed $Pc$ structure with constrained magnetic moments and SOC.}
\label{DOS1}
\end{figure}

\begin{table}[h]
  \centering
  \caption*{Table S2. Lattice parameters and atomic coordinates of the fully relaxed $P$2$_1/c$ structure calculated in $m\Gamma^+_1$ configuration of Mn spins in absence of SOC. Here, we use $U$ = 2.5 eV and $J$ = 0.5 eV for the Mn-$d$ orbitals.}
  \label{tab-P21c}
  \begin{tabular}{|c|c|c|c|c|c|}
    \hline
    Cell parameters& Fully relaxed values& Atomic sites& x &y &z\\
    \hline\rule{0pt}{1.0\normalbaselineskip}
     $a$ (\AA)&7.15412&Ba1&-0.00255&0.68419& -0.03411\\
     &&Ba2&0.34559 & 0.65133 &0.69502\\
     $b$ (\AA)& 10.01275&Ba3&0.50000 & 0.00000& 0.50000 \\
     &&Ba4&0.17911 & 0.50001& 0.33705\\
     $c$ (\AA)&10.65874&Mn1&0.57502 & 0.66605& 0.42500 \\
     & &Mn2&0.76977 & 0.66933& 0.22403\\
     $\alpha$ (\degree) & 90.00000&O1&0.51368 & 0.60016& 0.25750\\
      & &O2&0.67153  &0.68318& 0.83624\\
     $\beta$ (\degree) & 92.01370 &O3&0.34344 & 0.74028& 0.44629\\
    & &O4&0.82562 & 0.58837 &0.38620\\
     $\gamma$ (\degree)&90.00000&O5&0.67286 & 0.73098 &0.58202\\
     &&O6&0.50000 & 0.00000 &0.00000\\
     $V_{\rm{cell}}$ ({\AA}$^3$)&763.04046&O7&0.83641 & 0.52249 &0.13298\\
     &&O8&0.00198 & 0.74910 &0.21381\\
    \hline
\end{tabular}
\end{table}

\begin{figure}[h]
\centering
\includegraphics[scale=1.3]{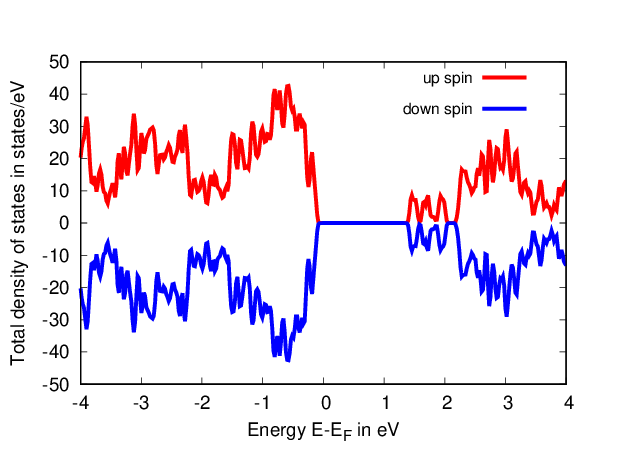} 
\caption*{Figure S2. Total density of states calculated in the $m\Gamma^{+}_1$ configuration with $U$ = 2.5 eV and $J$ = 0.5 eV in absence of SOC.}
\label{DOS2}
\end{figure}

  \begin{table}[h]
  \centering
  \caption*{Table S3. Summary of the self-consistent non-collinear DFT calculations including SOC for $U$ = 2.5 eV and $J$ = 0.5 eV. Here, $\theta$ denotes the angle between the magnetic moments of the Mn atoms within individual Mn$_2$O$_9$ dimers. $F$ is the free energy of a particular configuration with respect to the ground state free energy. Fig.~\ref{Pc-nc} of the main text illustrates the different initial and final non-collinear spin configurations.}
  \label{tab-nc1}
  \begin{tabular}{|c|c|c c c|c|c|c|}
    \hline
    Config.&Inequivalent &\multicolumn{3}{c|}{Final spin moments ($\mu_B$)}& $\theta$ ($\degree$)   &$F$  & Final \\
     & Mn sites&$m_x$&$m_y$&$m_z$& & (meV)&s.g. \\
     \hline
     &Mn1&0.01100 & 2.29300 & 1.53500 &$\theta_{\text{Mn1}-\text{Mn3}}$&&\\
    AFM1 &Mn2&0.01800 & 2.24250 & 1.61250 &= 179.80&0.06482 &$P\bar{1}^\prime$\\
     &Mn3&-0.01300 & -2.29000& -1.54450 &$\theta_{\text{Mn2}-\text{Mn4}}$&&\\
     &Mn4&-0.02100 &-2.23400 &-1.62000 &= 179.76&&\\
    \hline
     &Mn1&-0.00700& 2.75900 & -0.05300 &$\theta_{\text{Mn1}-\text{Mn3}}$&&\\
    AFM2 &Mn2&0.00100 & 2.76200 & 0.04100 &= 179.77&0.00066 &$P\bar{1}^\prime$\\
     &Mn3&0.00500 & -2.76200 & 0.04200 &$\theta_{\text{Mn2}-\text{Mn4}}$&&\\
     &Mn4&-0.00300& -2.75900& -0.05200 &= 179.77&&\\
    \hline
    &Mn1&-0.02100 &2.23350 & -1.62050 &$\theta_{\text{Mn1}-\text{Mn3}}$&&\\
    AFM3 &Mn2&-0.01500& 2.29100 & -1.54250 &= 179.74&0.06832 &$P\bar{1}^\prime$\\
     &Mn3&0.02000 & -2.24300 & 1.61200 &$\theta_{\text{Mn2}-\text{Mn4}}$&&\\
     &Mn4&0.01450 & -2.29600 & 1.53100 &= 179.74&&\\
    \hline
     &Mn1&1.62000 &2.23400 & -0.02000 &$\theta_{\text{Mn1}-\text{Mn3}}$&&\\
    AFM4 &Mn2&1.54300 &2.29100 & -0.01300 &= 179.80&0.06280 &$P\bar{1}^\prime$\\
     &Mn3&-1.61150 & -2.24250 & 0.01750 &$\theta_{\text{Mn2}-\text{Mn4}}$&&\\
     &Mn4&-1.53300 & -2.29500 &0.01150 &= 179.80&&\\
    \hline
    &Mn1&-0.05225& 2.75900 & 0.00500 &$\theta_{\text{Mn1}-\text{Mn2}}$&&\\
    AFM5 &Mn2&0.04125 & -2.76200 & -0.00300 &= 179.77&0.00051 &$P2^\prime_1/c$\\
    \hline
    &Mn1&-1.52900 &2.29750 & 0.01850 &$\theta_{\text{Mn1}-\text{Mn3}}$&&\\
    AFM6 &Mn2&-1.60700 &2.24600 & 0.02400 &= 179.80&0.06280 &$P\bar{1}^\prime$\\
     &Mn3&1.53800 & -2.29400 & -0.01950 &$\theta_{\text{Mn2}-\text{Mn4}}$&&\\
     &Mn4&1.61350 & -2.23900 &-0.02650 &= 179.80&&\\
    \hline
    &Mn1&2.26750 & -0.03025& 1.57250 &$\theta_{\text{Mn1}-\text{Mn2}}$&&\\
    AFM7 &Mn2&-2.27050 &0.02600 & -1.57250 &= 179.90& 0.03023 &$P2_1/c^\prime$\\
    \hline
    &Mn1&2.76050 & -0.00250 &-0.00750 &$\theta_{\text{Mn1}-\text{Mn2}}$&&\\
    AFM8 &Mn2&-2.76200& 0.00225 & 0.00725 &= 180.00& 0.14096 &$P2_1/c^\prime$\\
    \hline
    &Mn1&2.25150 & -0.02700 & -1.59600 &$\theta_{\text{Mn1}-\text{Mn2}}$&&\\
    AFM9 &Mn2&-2.25450 &-0.02225& 1.59550 &= 179.90& 0.28561 &$P2_1/c^\prime$\\
    \hline
\end{tabular}
\end{table}

  \begin{table}[h]
  \centering
  \caption*{Table S4. Summary of the self-consistent non-collinear DFT calculations including SOC for $U$ = 2.5 eV and $J$ = 0.5 eV. Here, $\theta$ denotes the angle between the magnetic moments of the Mn atoms within individual Mn$_2$O$_9$ dimers. $F$ is the free energy of a particular configuration with respect to the ground state free energy. Fig.~\ref{Pc'-nc} of the main text illustrates the different initial and final non-collinear spin configurations.}
  \label{tab-nc2}
  \begin{tabular}{|c|c|c c c|c|c|c|}
    \hline
    Config.&Inequivalent &\multicolumn{3}{c|}{Final spin moments ($\mu_B$)}& $\theta$ ($\degree$)   &$F$  & Final \\
     & Mn sites&$m_x$&$m_y$&$m_z$& & (meV)&s.g. \\
    \hline
    &Mn1&0.04300 & -1.57200 & 2.26800 &$\theta_{\text{Mn1}-\text{Mn3}}$&&\\
    AFM10 &Mn2&0.03700 & -1.65000 &2.21500 &= 179.80&0.12668 &$P\bar{1}^\prime$\\
     &Mn3&-0.04050 &1.58100 & -2.26400 &$\theta_{\text{Mn2}-\text{Mn4}}$&&\\
     &Mn4&-0.03600 &1.65600 & -2.20750 &= 179.80&&\\
    \hline    
     &Mn1&0.18000 & 0.05150 & 2.75300 &$\theta_{\text{Mn1}-\text{Mn2}}$&&\\
    AFM11 &Mn2&-0.18000 &0.04325& -2.75600 &= 179.83&0.18315  &$P2_1/c^\prime$\\
    \hline
    &Mn1&0.03500 & 1.63900 & 2.22000 &$\theta_{\text{Mn1}-\text{Mn3}}$&&\\
    AFM12 &Mn2&0.03900 & 1.56400 & 2.27600 &= 179.83&0.12834 &$P\bar{1}^\prime$\\
     &Mn3&-0.03600 &-1.63400& -2.22700 &$\theta_{\text{Mn2}-\text{Mn4}}$&&\\
     &Mn4&-0.04100 &-1.55500 &-2.27900 &= 179.80&&\\
    \hline
\end{tabular}
\end{table}

\begin{table}[h]
  \centering
  \caption*{Table S5. Summary of the self-consistent collinear calculations including SOC for $U$ = 2.5 eV and $J$ = 0.5 eV. Here, $\theta$ denotes the angle between the magnetic moments of the Mn atoms within individual Mn$_2$O$_9$ dimers. $F$ is the free energy of a particular configuration with respect to the ground state free energy. Fig.~\ref{coll} of the main text illustrates the different initial collinear spin configurations.}
  \label{tab-coll}
  \begin{tabular}{|c|c|c c c|c|c|c|}
    \hline
    Config.&Inequivalent &\multicolumn{3}{c|}{Final spin moments ($\mu_B$)}& $\theta$ ($\degree$)  &$F$ & Final \\
     & Mn sites&$m_x$&$m_y$&$m_z$& &(meV)  &s.g. \\
     \hline
    &Mn1&2.76000 & -0.00175 &-0.00150 &$\theta_{\text{Mn1}-\text{Mn2}}$&&\\
    AFM13 &Mn2&-2.76200 &0.00125 &0.00150 &= 180.00&0.13961 &$P2_1/c^\prime$\\
    \hline
    
    &Mn1&-2.78200 &-0.00375 &-0.00475 &$\theta_{\text{Mn1}-\text{Mn2}}$&&\\
    AFM14 &Mn2&2.76600 & 0.00500 & -0.00150 &= 179.87&209.10613 &$P$2$_1/c$\\
    \hline
    
    &Mn1&0.00500 & 2.75900 & 0.05200 &$\theta_{\text{Mn1}-\text{Mn2}}$&&\\
    AFM15 &Mn2&-0.00300 &-2.76200 &-0.04100&= 179.77&0.00059 &$P2^\prime_1/c$\\
    \hline
    
    &Mn1&-0.01500& -2.78100 &-0.08750 &$\theta_{\text{Mn1}-\text{Mn2}}$&&\\
    AFM16 &Mn2&0.01350 & 2.76500  &0.08175&= 179.90&209.06489 &$P2^\prime_1/c^\prime$\\
    \hline
    
    &Mn1&0.02300 & -0.05100 & 2.75900 &$\theta_{\text{Mn1}-\text{Mn2}}$&&\\
    AFM17 &Mn2&-0.02200& 0.04300 & -2.76200&= 179.83&0.19357 &$P2_1/c^\prime$\\
    \hline
    
    &Mn1&0.44400 & 2.69500 & 0.39500 &$\theta_{\text{Mn1}-\text{Mn3}}$&&\\
    AFM18 &Mn2&0.45200 & 2.68100 & 0.48600 &= 179.80&0.00000 &$P\bar{1}^\prime$\\
     &Mn3&-0.44650& -2.69500 &-0.4050 &$\theta_{\text{Mn2}-\text{Mn4}}$&&\\
     &Mn4&-0.45350& -2.67700 &-0.49500 &= 179.80&&\\
    \hline
\end{tabular}
\end{table}

\begin{table}[h]
  \centering
  \caption*{Table S6. Ground state magnetic configurations obtained for different sets of $U$ and $J$ values. Here, $\theta$ denotes the angle between the magnetic moments of the Mn atoms within individual Mn$_2$O$_9$ dimers.}
  \label{tab-U}
  \begin{tabular}{|c|c|c c c|c|c|}
    \hline
    Values of &Inequivalent &\multicolumn{3}{c|}{Final spin moments ($\mu_B$)}& $\theta$ ($\degree$) & Final \\
     $U$ and $J$ (eV)& Mn sites&$m_x$&$m_y$&$m_z$& &space group \\
     \hline
    &Mn1&0.00000 & 2.54400 & -0.03500 &$\theta_{\text{Mn1}-\text{Mn3}}$&\\
    $U = 0.0$ &Mn2&0.00200 & 2.55700 & 0.03400 &= 179.86 &$P\bar{1}^\prime$\\
     $J = 0.0$&Mn3&0.00000 & -2.55800 &0.02900 &$\theta_{\text{Mn2}-\text{Mn4}}$&\\
     &Mn4&-0.00200 &-2.54400 &-0.04100 &= 179.84&\\
    \hline
    
    &Mn1&-0.44400 & 2.69500 & 0.39500 &$\theta_{\text{Mn1}-\text{Mn3}}$&\\
    $U = 2.5$ &Mn2&0.45200 & 2.68100 & 0.48600 &= 179.80 &$P\bar{1}^\prime$\\
     $J = 0.5$&Mn3&-0.44650& -2.69500 &-0.4050 &$\theta_{\text{Mn2}-\text{Mn4}}$&\\
     &Mn4&-0.45350& -2.67700 &-0.49500 &= 179.80&\\
    \hline

    &Mn1&0.46950 & 2.89500&  0.40800 &$\theta_{\text{Mn1}-\text{Mn3}}$&\\
    $U = 5.0$ &Mn2&0.48600 & 2.87100 & 0.53600 &= 179.70 &$P\bar{1}^\prime$\\
     $J = 1.0$&Mn3&-0.47200 &-2.89250 &-0.42250 &$\theta_{\text{Mn2}-\text{Mn4}}$&\\
     &Mn4&-0.48900& -2.86900 &-0.55000 &= 179.72&\\
    \hline
    
     &Mn1&0.51100 & 3.24350 & 0.41650 &$\theta_{\text{Mn1}-\text{Mn3}}$&\\
    $U = 8.0$ &Mn2&0.54700&  3.20000 & 0.64750 &= 179.75 &$P\bar{1}^\prime$\\
     $J = 1.0$&Mn3&-0.51300 &-3.24200 &-0.43100 &$\theta_{\text{Mn2}-\text{Mn4}}$&\\
     &Mn4&-0.54900& -3.19600& -0.66100 &= 179.75&\\
    \hline
\end{tabular}
\end{table}

\begin{table}[h]
  \centering
  \caption*{Table S7. Components of the static Dielectric permittivity tensor $\epsilon^\infty_{ij} ~(i, j = x, y, z)$ calculated for the relaxed $P$2$_1/c$ structure in the $m\Gamma^+_1$  magnetic configuration without SOC using $U = 2.5$ eV and $J = 0.5$ eV. $\epsilon^{\infty}_{\text{el}}$ and $\epsilon^{\infty}_{\text{ion}}$ represent the electronic and ionic contributions, respectively.}
  \label{tab-diel}
  \begin{tabular}{|c c c c c c c c c c|}
    \hline
     $\epsilon^{\infty}$&$xx$&$xy$&$xz$&$yx$&$yy$&$yz$&$zx$&$zy$&$zz$ \\
    \hline
    $\epsilon^{\infty}_{\text{el}}$&5.543&0.000&-0.291&0.000&5.435&0.000&-0.291&0.000&6.085\\[0.1cm]
    $\epsilon^{\infty}_{\text{ion}}$&11.891&0.000&0.452&0.000000&14.512&0.000&0.452&0.000&18.006\\[0.1cm]
    \hline
\end{tabular}
\end{table}

\begin{table}[h]
  \centering
  \caption*{Table S8. Components of the Born effective charge tensors $Z^*_{ij} ~(i, j = x, y, z)$ in units of $e$ calculated for the relaxed $P$2$_1/c$ structure in the $m\Gamma^+_1$  magnetic configuration without SOC using $U = 2.5$ eV and $J = 0.5$ eV. Positions of the symmetry-unique atomic sites are given in Table~S2.}
  \label{tab-BEC}
  \begin{tabular}{|c c c c c c c c c c|}
    \hline
    Sites&$Z^*_{xx}$&$Z^*_{xy}$&$Z^*_{xz}$&$Z^*_{yx}$&$Z^*_{yy}$&$Z^*_{yz}$&$Z^*_{zx}$&$Z^*_{zy}$&$Z^*_{zz}$ \\
    \hline
    Ba1&3.362&-0.378&-0.382&-0.425&2.292&-0.091&-0.367&-0.150&2.964\\[0.08cm]
    Ba2&2.808&0.212&-0.233&0.252&2.844&0.339&-0.197&0.330&2.544\\[0.08cm]
    Ba3&2.196&-0.631&0.485&-0.718&3.671&-0.008&0.646&0.011&3.267\\[0.08cm]
    Ba4&3.361&-0.332&0.370&-0.280&3.140&-0.021&0.432&0.030&2.317\\[0.08cm]
    Mn1&3.851&0.478&-0.615&0.711&4.187&0.078&-0.664&-0.132&5.352\\[0.08cm]
    Mn2&3.519&0.214&-0.539&0.073&3.255&0.517&-0.625&0.735&4.925\\[0.08cm]
    O1&-2.346&-0.452&0.083&-0.599&-1.904&-0.274&0.204&-0.353&-2.885\\[0.08cm]
    O2&-1.910&0.056&0.359&0.124&-2.804&-0.224&0.623&-0.257&-2.456\\[0.08cm]
    O3&-3.305&0.644&0.336&0.599&-2.020&-0.017&0.205&0.042&-1.776\\[0.08cm]
    O4&-2.598&0.468&0.024&0.382&-1.853&0.425&0.154&0.3753&-2.719\\[0.08cm]
    O5&-1.772&-0.469&-0.613&-0.706&-2.127&-1.333&-1.016&-1.478&-4.366\\[0.08cm]
    O6&-1.789&1.347&0.558&1.084&-4.382&-1.230&0.555&-1.319&-2.470\\[0.08cm]
    O7&-1.865&0.575&0.519&0.479&-2.869&-0.763&0.461&-0.683&-2.307\\[0.08cm]
    O8&-3.308&-0.734&0.170&-0.706&-1.785&0.162&0.189&0.144&-1.992\\[0.08cm]
    \hline
\end{tabular}
\end{table}

\end{document}